\documentclass[twocolumn,showpacs,preprintnumbers,amsmath,amssymb]{revtex4}
\preprint{}
\usepackage{graphicx}% Include figure files
\usepackage{dcolumn}% Align table columns on decimal point
\usepackage{bm}% bold math
\usepackage{mathrsfs}
\begin{document}
\title{Quantum Metrology in Correlated Environments}
\author{Dong  Xie}
\email{xiedong@mail.ustc.edu.cn}
\author{An Min Wang}
 \email{anmwang@ustc.edu.cn}
  \affiliation{Department of Modern Physics , University of Science and Technology of China, Hefei, Anhui, China.}
\begin{abstract}
We analytically obtain the precision bounds of frequency measurements in
correlated Markovian and non-Markovian environments by using a variational
approach. It is verified that in standard Ramsey spectroscopy setup, the metrological equivalence of product and
maximally entangled states persists in
maximally correlated Markovian and non-Markovian environments. We find that
the optimal measurement can achieve a much higher resolution than standard
Ramsey spectroscopy in the correlated environments. When the number of
particles in the maximally entangled states is even, the precision bound
decreases with interrogation time; and when the number is odd, the precision
bound is independent of interrogation time, both in correlated Markovian
and general non-Markovian environments. In addition, the opposite case can
appear in some special non-Markovian environments.
\end{abstract}
\maketitle
\section{Introduction}
 Quantum metrology is a fundamental and important subject, which concerns the estimation of parameters under the constraints of quantum dynamics \cite{lab1,lab2,lab3,lab4, lab5}. The Cram$\acute{e}$r-Rao bound limits the uncertainty in the estimation of a parameter \cite{lab6,lab7,lab8}.

 Environments can bring great impact on quantum systems, leading to play a very important role in quantum metrology.  Without suffering from environments, entangled states can achieve a higher resolution as compared to the precision limits achievable with uncorrelated probes \cite{lab9,lab10}. In the real experiments, environments will induce decoherence, which can affect the measurement precision. S. F. Huelga et al. \cite{lab11} first studied precision spectroscopy in the presence of Markovian dephasing, and showed that given a fixed number of particles $n$ and a total available time $T$, uncorrelated and maximally entangled particles can achieve exactly the same precision when subject to Markovian dephasing. Recently, Yuichiro Matsuzaki et al. \cite{lab12} and  Alex W. Chin et al. \cite{lab13} explored quantum metrology in non-Markovian environments respectively, and achieved that the metrological equivalence didn't hold.

 Correlations between the environments can contribute to the increase of the quantum efficiency of transport \cite{lab14}.  In photosynthetic light harvesting, environmental correlation effects help the excitation energy transfer\cite{lab15}. Correlated environments can generate strong nonlocal memory effects, although the local dynamics is Markovian \cite{lab16}. Role of environmental correlations has also been investigated in the non-Markovian dynamics of a spin chain system\cite{lab17}.
 In this article, we analyze whether the metrological equivalence persists when the whole system suffers from correlated Markovian and non-Markovian environments, which are unexplored up to now.
 In fact, correlated environments maybe make the entangled probes avoid or slow the global dephasing rate, leading to that correlated environments have advantage over uncorrelated environments in quantum metrology. We mainly research precision bounds when correlated environments keep or increase the dephasing rate of the whole system. Namely, what's the best precision, which is achieved by the optimal measurement in correlated environments.

 The variational approach \cite{lab18} and some symmetries are used to research the precision bounds
and solve the open question left by Ref. \cite{lab13}. For solving the problem with respect to the correlated environments, we define a general function $F(w)$, which can effectively express the impact of correlations among environments on the dephasing rate.  Significantly and interestingly, it is found that the standard Ramsey spectroscopy isn't optimal in the correlated environments, and the optimal measurement can achieve a much lower frequency uncertainty. If the probes are in product state, the precision bound decreases with the interrogation time $t$ when the number of particles $n$ is even. So if experiments allow, the precision bound can be close to 0 when the interrogation time $t$ is very large. And when the number of particles $n$ is odd, the precision bound is independent of the interrogation time $t$. The same situation exists when the probes are prepared in maximally entangled state. It reflects that certain symmetries play an important role. It's worth mentioning that in some special non-Markovian environments, the opposite case appears. What's more, comparing with the case in the uncorrelated environments, the correlations help to obtain a better resolution by using the optimal measurement.

 The rest of this article is arranged as follows. In section II, we study the precision bounds in uncorrelated Markovian and non-Markovian environments with the help of variational approach. The precision bounds in correlated  Markovian and non-Markovian environments are mainly explored in the section III. In section IV, we discuss about the definition of Markovianity and non-Markovianity, the influence of  odd and even particles on the precision, and the maximally correlated environments in experiment.  Finally,
we draw our conclusion in section V.
\section{uncorrelated environments}
Let us consider a global system composed of $n$ particles. The Hamiltonian of each system is described by $w_0Z$ ($\hbar=1$ in the whole article). The $n$ particles suffer from the corresponding $n$ uncorrelated environments, which induce the pure dephasing. The eigenvector of Pauli operator $Z$ is denoted by $(|0\rangle,|1\rangle)$. The time evolution of the reduced density matrix of the system (for one particle) is given by
\begin{equation}
\rho_{ii}(t)=\rho_{ii}(0),
\end{equation}
\begin{equation}
\rho_{01}(t)=\rho_{01}(0)e^{-2\gamma(t)},
\end{equation}
for $i$=0, 1.

When the environment induces a pure Markovian dephasing, the function $\gamma(t)=\gamma t$ ($\gamma$ is the decay rate). Ramsey spectroscopy \cite{lab19} gets the same frequency resolution for maximally entangled and product state (using the same notation as in Ref. \cite{lab11,lab13})
\begin{equation}
|\delta{w_0|_e}|=|\delta{w_0|_u}|=\sqrt{\frac{2 e\gamma}{nT}},
\end{equation}
where $T$ denotes the total duration of the experiment, and $w_0$ is the the atomic frequency.

The optimal frequency resolution from the best measurement can be obtained by the variational approach in Ref. \cite{lab16}. The quantum Fisher information (QFI) is given by
\begin{equation}
\mathcal {F}_{\textit{Q}}[\hat{\rho}_S(\phi)]=\min_{\substack{\hat{h}_E(\phi)}}4\langle[ \mathcal{\hat{H}(\phi)}-\langle\mathcal{\hat{H}(\phi)}\rangle]^2\rangle_\Phi,
\end{equation}
in which, $\mathcal{\hat{H}(\phi)}=\hat{H}_{S,E}(\phi)-\hat{h}_E(\phi)$, $\phi$ is the detuning between the frequency $w$ of the external oscillator and the atomic frequency $w_0$ to which we intend to lock it to, $|\Phi_{S,E}(\phi)\rangle$ is a purification of $\hat{\rho}_S(\phi)$, and $\hat{h}_E$ is the Hermitian operator in the space of environment.
The best resolution is described by the expression
\begin{equation}
\delta w_0^2=\frac{1}{N\mathcal {F}_{\textit{Q}}[\hat{\rho}_S(\phi)]},
\end{equation}
where the total number of experiment data $N=nT/t$.

According to the principle of variational approach in Ref. \cite{lab18} (see the Appendix), we choose the state
\begin{equation}
|\Phi_{S,E}(\phi)\rangle=\prod_{i=1}^ne^{-i\phi t Z_i/2}e^{-i\arccos(\sqrt{P(\gamma t)})Z_iY_i^\textmd{E}}|\psi\rangle|0\rangle^{\otimes n}_E,
\end{equation}
where $P(\gamma t)=\frac{1+\exp(-\gamma t)}{2}$; $Z_i$, $Y_i^E$ are Pauli operators for the $i$th system and environment respectively; $|\psi\rangle$ denotes the initial state of the whole system. Suppose that $n$ systems are completely identical, and the corresponding $n$ environments are also same. Based on the symmetry, let the operator
\begin{equation}
\hat{h}_E(\phi)=\sum_{i=1}^n\alpha X_i^E+\beta Y_i^E+\delta Z_i^E,
\end{equation}
where $\alpha$, $\beta$, and $\delta$ are three variational parameters. Then, substituting Eq.(6) and Eq.(7) into Eq.(4), we can obtain the minimum value by taking the derivative of three variational parameters above. As a result, the resolution is given by
\begin{equation}
\delta w_0^2(t)=(1-\langle\sum_{i=1}^nZ_i/n \rangle^2_\psi)\frac{1+nq[\exp(2\gamma t)-1]}{qn^2Tt},
\end{equation}
 where $q=\frac{\Delta(\sum_{i=1}^nZ_i/n)^2}{1-\langle\sum_{i=1}^nZ_i/n \rangle^2_\psi}$\cite{lab20}.
 The best resolution $|\delta w_0|_{\textmd{opt}}=\sqrt{\frac{2\gamma}{nT}}$, when $q=1$ and $ \langle\sum_{i=1}^nZ_i/n \rangle^2_\psi=0$.

 Next, we consider that environments induce the pure non-Markovian dephasing. As shown in Ref. \cite{lab13}, we also study the simple power law form of $\gamma(t)=\gamma t^\nu$. The case of $\nu=1$ corresponds to the Markovian case. And it's worth noting that the case of $\nu=2$ isn't a specific feature of chosen model, but rather a general consequence of the unitary evolution of the total system and environment state.

 Using the Ramsey spectroscopy setup, the resolution for an initial preparation of $n$ particles in a product state $(\frac{|0\rangle+|1\rangle}{\sqrt{2}})^{\otimes n}$ and a maximally entangled state $\frac{|0\rangle^{\otimes n}+|1\rangle^{\otimes n}}{\sqrt{2}}$ are given by
 \begin{equation}
|\delta{w_0|_u}|^R=\sqrt{\frac{(2 e \gamma\nu)^{1/\nu}}{nT}},
\end{equation}
  \begin{equation}
|\delta{w_0|_e}|^R=\sqrt{\frac{(2 e \gamma\nu)^{1/\nu}}{n^{(2-1/\nu)}T}}.
\end{equation}

 For the non-Markovian case, the function $P(\gamma t)=\frac{1+\exp(-\gamma t^\nu)}{2}$ in the Eq.(6). So the optimal measurement achieves that the resolution is similar to Eq.(8)
 \begin{equation}
|\delta w_0(t)|=\sqrt{(1-\langle\sum_{i=1}^nZ_i/n \rangle^2_\psi)\frac{1+nq(\exp(2\gamma t^\nu)-1)}{qn^2Tt}}.
\end{equation}
As a result, the optimal resolution is
\begin{equation}
|\delta w_0|_{\textmd{opt}}=\sqrt{\frac{(2\gamma \nu)^{1/\nu}}{(1-\frac{1}{2\nu})^{(1-1/\nu)}n^{(2-1/\nu)}T}}
\end{equation}
for $\nu\geq1$ and $n\gg1$, when $q=1$ and $ \langle\sum_{i=1}^nZ_i/n \rangle^2_\psi=0$.

 Then we define that the improvement
 \begin{equation}
 I=\frac{\min\{|\delta{w_0|_e}|^R,|\delta{w_0|_u}|^R\}}{|\delta w_0|_{\textmd{opt}}}.
 \end{equation}
 For $\nu>1$, the maximum improvement of $I=\frac{|\delta{w_0|_e}|^R}{|\delta w_0|_{\textmd{opt}}}=[\frac{e}{(1-\frac{1}{2v})^{(1-\nu)}}]^{\frac{1}{2\nu}{}}$ in the resolution is achievable. When $\nu=1$ (for the Markovian case), the maximum value of the improvement $I=\sqrt{e}$, which is known result. As shown in Fig.1, for $\nu>1$ the improvement is a mere constant to be of the order of 1. For the most general non-Markovian dephasing ($\nu=2$), the improvement $I=1.2$. For $0<\nu<1$, the analytical solution is hard to get, but it is easy to verify it numerically in Eq.(11). For example, when $\gamma=1/2$, $n=100$ and $\nu=1/4$, the improvement $I=1$. We perform a lot of validation, and find that the improvement $I$ is of the  order of 1. In a word, it follows the guess in Ref. \cite{lab13} that like the Markovian case ($\nu=1$), the improvement $I$ is of the order of 1 in the non-Markovian environment.
\begin{figure}[h]
\includegraphics[scale=1.4]{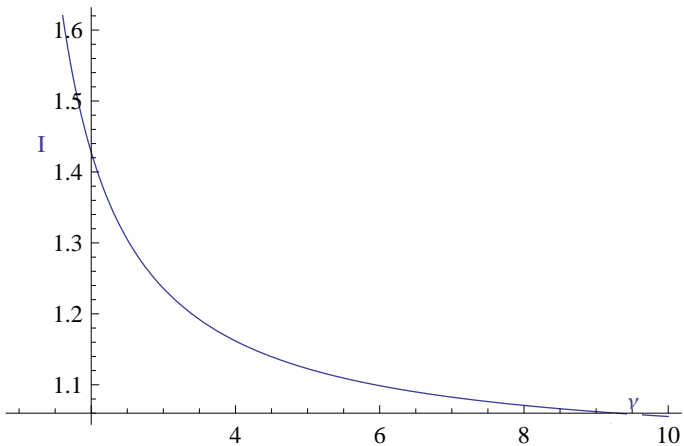}
 \caption{\label{fig.1}The graph shows that the improvement $I$ changes with the power $\nu$, for $\nu\geq1$.    When $\nu$ approaches infinity, the improvement $I$ is close to 1. And when $\nu=1$, the improvement $I=\sqrt{e}$ corresponding to the Markovian case. }
 \end{figure}
\section{Correlated Environments}
In quantum metrology, the roles of correlated environments are unexplored up to now. It's trivial that the correlations among environments can make the entangled states of whole system avoid decoherence, leading to the improvement of resolution. Nontrivial question arises: what is the optimal resolution when the correlations keep or accelerate the global dephasing rate of whole system.

The pure dephasing is induced by the system-environment interaction. The coupling to environment is described by $Z\otimes \hat{B}$, where $\hat{B}$ is operator of environment. So the dephasing of a system is given by
\begin{equation}
\exp[-\gamma(t)]=\textmd{Tr}[\hat{\rho}_E(0)e^{-i(\hat{B}+\hat{H}_E)t}e^{-i(\hat{B}-\hat{H}_E)t]},
\end{equation}
where $\hat{H}_E$ represents the Hamiltonian of an environment and $\hat{\rho}_E(0)$ denotes the initial density matrix of an environment.
Then we can use a general function $F(w)$ to express the above equation
\begin{equation}
\exp[-\gamma(t)]=\int_{-\infty}^{\infty}dwF(w)\exp(-iwt),
\end{equation}
where the general function $F(w)\geq0$ and $\int_{-\infty}^{\infty}dwF(w)=1$.
 For example, the bath is composed of harmonic oscillators. The general function $F(w)$ is described by a spectral function $J(w)$\cite{lab21,lab22,lab23}
\begin{equation}
\begin{split}
F(w)=&1/(2\pi)\int_{-\infty}^\infty dt\exp[1/2\int_0^\infty\\
&dw'J(w')\coth(w'\beta/2)\frac{1-\cos(w't)}{w'^2}]\exp(iwt),
\end{split}
\end{equation}
where $\beta$ is the inverse of temperature.

For $n$ same environments, the general function
\begin{equation}
\begin{split}
F(w_1,...,w_n)=&\sin^2\theta F(w_1)F(w_2)...F(w_n)+\\
&\cos^2\theta F(w_1)\delta(w_1-w_2)...\delta(w_1-w_n).
\end{split}
\end{equation}
When $\sin\theta=1$, the environments are uncorrelated; when $\sin\theta=0$, the environments are maximally correlated. And if $n$ particles is subject to a single environment, it can also be treated as $n$ maximally correlated environments.
For $n$ particles, the total dephasing function of $(|0\rangle\langle1|)^{\otimes n}$ in maximally correlated environments is described by
\begin{equation}
\begin{split}
\exp[-\Gamma(t)]=\frac{\textmd{Tr}[(|0\rangle\langle1|)^{\otimes n}\rho_S(t)]}{\textmd{Tr}[(|0\rangle\langle1|)^{\otimes n}\rho_S(0)]}\\
=\int_{-\infty}^{\infty}dwF(w)\exp(-inwt).
\end{split}
\end{equation} For a single particle, the dephasing function $\exp[-\gamma(t)]=\int_{-\infty}^{\infty}dwF(w)\exp(-iwt)$. Then we can obtain that $\Gamma(t)=\gamma (nt)^\nu$, according to that $\gamma(t)=\gamma t^\nu$.

In the maximally correlated environments, the frequency resolution is achieved by Ramsey spectroscopy: for the maximally entangled state $\frac{|0\rangle^{\otimes n}+|1\rangle^{\otimes n}}{\sqrt{2}}$
\begin{equation}
\delta{w_0}^2|_e=\frac{1}{n^2Tt_e}e^{2n^\nu\gamma t_e^\nu},
\end{equation}
for the product state $(\frac{|0\rangle+|1\rangle}{\sqrt{2}})^{\otimes n}$
\begin{equation}
\delta{w_0}^2|_u=\frac{1}{nTt_u}e^{2\gamma t_u^\nu}.
\end{equation}
The optimal interrogation time $t_e=(\frac{1}{2n\gamma\nu})^{(1/\nu)}$ and $t_u=(\frac{1}{2\gamma\nu})^{(1/\nu)}$. So the relative
frequency resolution of entangled and uncorrelated probes $r=\frac{|\delta{w_0}|_u}{|\delta{w_0}|_e}=1$. Namely, in maximally correlated environments, these two initial preparations of probes are metrologically equivalent for the standard Ramsey spectroscopy.

The optimal measurement can improve the resolution greatly. Firstly, we treat the maximally correlated environments as a single environment.
Choose the purification of $\hat{\rho}_S(\phi)$
\begin{equation}
|\Phi_{S,E}(\phi)\rangle=e^{-i\phi tZ/2}e^{-i\arccos(\sqrt{P(\gamma t)})|Z|^{(\nu-1)}ZY^\textmd{E}}|\psi\rangle|0\rangle,
\end{equation}
where $Z=\sum_{i=1}^nZ_i$ and $P(\gamma t)=\frac{1+\exp(-\gamma t^\nu)}{2}$.
And the operator
\begin{equation}
\hat{h}_E(\phi)=\alpha X^E+\beta Y^E+\delta Z^E.
\end{equation}
Use the variational approach to get the QFI and achieve the optimal resolution.
For the general power law dependence on time $\gamma(t)=\gamma t^\nu$, we obtain
\begin{equation}|\delta w_0|=
\begin{cases} \frac{1}{\sqrt{t(n-n^2\frac{(\sum_{i=0}^nC_n^i\sin(2|n-2i|^\nu\varphi))^2}{2^{2n}})}},&\textmd{ product state}; \\ \frac{1}{\sqrt{t(n^2-n^2\sin^2(2n^\nu\varphi))}},& \textmd{ entangled state};
\end{cases}
\end{equation}
where $\cos2\varphi=e^{-\gamma t^\nu}.$
From the above equation, for probes in the product state and in the maximally entangled state, the optimal interrogation time $t$ can be very large when $n^\nu$ is even (because when M is even, $\textmd{lim}_{t\rightarrow\infty}\sin(2M\varphi)\approx0$). When $n^\nu$ is odd, the optimal interrogation time t is finite. So for the Markovian dephasing ($\nu=1$) and non-Markovian dephasing (such as $\nu=2$), when the number $n$ is even the optimal resolution can be close to $0$. And for some special non-Markovian dephasing, when $n$ is odd the optimal resolution can be close to 0 (because for some $\nu$, $n^{\nu}$ is even when $n$ is odd). So comparing with Ramsey spectroscopy, when $n^\nu$ is even, the optimal  measurement can improve the resolution greatly. The difference between even and odd reflects some symmetries, because we consider that all systems are same, and all environments are also completely identical.

Then we consider that correlations of environments aren't maximal. Suppose that the initial environments are in the pure state
\begin{equation}
|\Psi\rangle_E=A\frac{|0\rangle^{\otimes n}+|1\rangle^{\otimes n}}{\sqrt{2}}+B(\frac{|0\rangle+|1\rangle}{\sqrt{2}})^{\otimes n},
\end{equation}
where $A^2+B^2+2^{1/2-n/2}AB=1$ and the strength of correlations depends on the parameter $A$ (when $A=1$ the environments is maximally correlated).
Consider a purification of $\rho_S(\phi)$
\begin{equation}
|\Phi_{S,E}(\phi)\rangle=\prod_{i=1}^ne^{-i\phi t Z_i/2}e^{-i\arccos(\sqrt{P(\gamma t)})Z_iZ_i^\textmd{E}}|\psi\rangle|\Psi\rangle^{\otimes n}_E,
\end{equation}
where $P(\gamma t)=\frac{1+\exp(-\gamma t^\nu)}{2}$.

For the general correlated environments, the operator $\hat{h}_E$ involves many variational parameters. Here, we just compute a simple situation ($n=2$) in the Markovian case.
The operator $\hat{h}_E$ is given by
\begin{equation}
\begin{split}
\hat{h}_E=&\alpha(X_1+X_2)+\beta(Y_1+Y_2)+\delta(Z_1+Z_2)+\\
&\lambda_1X_1X_2+\lambda_2Y_1Y_2+\lambda_3Z_1Z_2+r_1(X_1Y_2+Y_1X_2)\\
&+r_2(X_1Z_2+Z_1X_2)+r_3(Y_1Z_2+Z_1Y_2),
\end{split}
\end{equation}
where $X_i$, $Y_i$, and $Z_i$ are the Pauli operators about environments.
By variational approach, we obtain the resolution when $t$ is very large,
\begin{equation}
|\delta w_0|\simeq\frac{1}{\sqrt{t(2-8B^2(A/\sqrt{2}+B/2)^2(1+q))}},
\end{equation}
here, $q=\langle\psi| Z_1Z_2|\psi\rangle$. It shows that when the environments are partly correlated ($A>0$), the resolution can also be close to 0 for even particles.
\section{Discussion}
For the preciseness, we emphasize that in this article, in the Markovian environment the dephasing rate $\gamma(t)=\gamma t$, and in the non-Markovian environment the dephasing rate $\gamma(t)$ isn't linear. And we only consider the simple power form $\gamma(t)=\gamma t^\nu$ for the non-Markovian case. However, the case of $\nu=2$ is a very general form arising from
early time unitary dynamics. So it is interesting to study the simple power form. The case of $\nu=2$ perhaps isn't a very general form for long time because this form is normally relevant, only up to the correlation time of the environment. For example, in the Ohmic bath, the Markovian result will recover \cite{lab13} at long time.  Hence, for other forms of $\gamma(t)$ during long time, it is worth further studying (it is out of content of this article).

There is striking difference between even and odd particles for the precision. Physically, it is the symmetry that plays an important role, because that $n$ particles are same and the corresponding environments are also identical. The even and odd will affect the symmetry so that the result is different. Mathematically, the odd and even number will have different function on the global dephasing rate.

Experimentally, one can put the probe particles in a single environment, such as the cavity. So the probe particles interact with the same environment. In another word, the single environment can be treated as the maximally correlated environments.

\section{Conclusion}
Precision frequency metrology in correlated Markovian and non-Markovian environments is studied. Firstly, we use a variational approach to obtain the optimal resolution in uncorrelated Markovian environments, recovering the known result. And use the variational approach to achieve the best resolution in uncorrelated non-Markovian environments. As a result, non-Markovian case is similar to Markovian one: the improvement $I$ is of the order of 1.
Then, in maximally correlated environments, Ramsey spectroscopy achieves same resolution for the initial probes in the product state and the maximally entangled state. Comparing with Ramsey spectroscopy, the optimal measurement can give much better resolution. Especially, when the number of particles $n$ is even, the resolution can be close to 0, in both the Markovian and general non-Markovian case. For some special non-Markovian case, the number of particles is odd to make the resolution become better. It shows some symmetries to create the difference between odd and even. Finally, we consider the partly correlated environments, and obtain that the optimal resolution can also be close to $0$ for even particles at a longer interrogation time.

This article will inspire further research about correlated environments. It is meaningful to study quantum metrology in more complex environments such as asymmetric environments, and consider the constraints from practical experiments. Utilizing the correlations among environments effectively will benefit the quantum information processing \cite{lab24} and quantum computation \cite{lab25}.

\section*{ACKNOWLEDGMENT}
This work was supported by the National Natural Science Foundation of China under Grant No. 10975125 and No. 11375168.
\section*{APPENDIX: Summary of Variational Approach}
The Fisher information is defined by $f(x)=\sum_k p_k(x)[d\ln[p_k(x)]/dx]^2$, where $p_k(x)$ is the probability of obtaining the set of experimental results $k$ for the parameter value $x$. And the QFI is given by the maximum of the Fisher information over all measurement strategies allowed by quantum physics:
\begin{equation}
\mathcal {F}_{\textit{Q}}[\hat{\rho}(x)]=\max_{\substack{\{\hat{E}_k\}}}f[\hat{\rho}(x);\{\hat{E}_k\}],
\end{equation}
where positive operator-valued measures $\{\hat{E_k}\}$ represents a specific measurement device.

If the probe state is pure, $\hat{\rho}(x)=|\psi(x)\rangle\langle\psi(x)|$, the correspondent expression of the QFI is
\begin{equation}
\mathcal {F}_{\textit{Q}}[\hat{\rho}(x)]=4[\frac{d\langle\psi(x)|}{dx}\frac{d|\psi(x)\rangle}{dx}-|\frac{d\langle\psi(x)|}{dx}|\psi(x)\rangle|^2].
\end{equation}

When the state $\hat{\rho}(x)$ is mixed, the simple analytical expression isn't available. However, it is always possible to enlarge the size of the original Hilbert space $S$ and build a pure state $|\Phi_{S,E}(x)\rangle\langle\Phi_{S,E}(x)|$ in the enlarged space $S+E$ that fulfills the condition $\textmd{Tr}_E[|\Phi_{S,E}(x)\rangle\langle\Phi_{S,E}(x)|]=\hat{\rho}_S(x)$. The state $|\hat{\Phi}_{S,E}(x)\rangle$ is a purification of the state $\hat{\rho}_S(x)$ of the system.

A physically motivated upper bound $C_Q[\hat{\Phi}_{S,E}(x)\rangle]$ of $\mathcal {F}_{\textit{Q}}[\hat{\rho}_S(x)]$ are given:
\begin{equation}
C_Q[\hat{\rho}_{S,E}(x)]=\mathcal {F}_{\textit{Q}}[\hat{\rho}_{S,E}(x)]\geq\mathcal {F}_{\textit{Q}}[\hat{\rho}_S(x)].
\end{equation}
The reason is that when a system and an environment are monitored together, the information acquired about unknown parameter cannot be smaller than the information obtained when only the system is measured.
So the QFI can be obtained
\begin{equation}
 {F}_{\textit{Q}}[\hat{\rho}_S(x)]= \min_{\substack{|\Phi_{S,E}(x)\rangle}}C_Q[\hat{\rho}_{S,E}(x)].
\end{equation}

There is always a unitary operator $\hat{u}_E(x)$ that connects two purifications $|\Psi_{S,E}(x)\rangle$ and $|\Phi_{S,E}(x)\rangle$: $|\Psi_{S,E}(x)\rangle=\hat{u}_E(x)|\Phi_{S,E}(x)\rangle$ for the same state $\hat{\rho}_S(x)$. So, given a purification $\Phi_{S,E}(x)$, the QFI can be found by minimizing $C_Q[\hat{u}_E(x)\hat{\rho}_{S,E}(x)\hat{u}_E^\dagger(x)]$ over all unitary operators $\hat{u}_E(x)$ on E space.

Then, to define two Hermitian operator $\hat{h}_E(x)$ and $\hat{H}_{S,E}(x)$ by
\begin{equation}
\begin{split}
&\hat{h}_E(x)=i\frac{d\hat{u}_E^\dagger(x)}{dx}\hat{u}_E(x),\\
&i\frac{d|\Phi_{S,E}(x)\rangle}{dx}=\hat{H}_{S,E}(x)|\Phi_{S,E}(x)\rangle.
\end{split}
\end{equation}

Using the definitions above, one can derive the Eq.(4). In order to minimize $C_Q$, the optimum Hermitian operator $\hat{h}_E^{(\textmd{opt})}(x)$ should satisfy the equation:
\begin{equation}
\begin{split}
&\hat{h}_E^{(\textmd{opt})}\hat{\rho}_E(x)+\hat{\rho}_E(x)\hat{h}_E^{(\textmd{opt})}=\\
&i\textmd{Tr}_S[\frac{d|\Phi_{S,E}\rangle}{dx}\langle\Phi_{S,E}|-|\Phi_{S,E}\rangle\frac{d\langle\Phi_{S,E}|}{dx}].
\end{split}
\end{equation}
According to the equation above, one may guess the approximation for $\hat{h}_E^{(\textmd{opt})}(x)$ that depends on the variational parameters. In this article, at the same time we utilize the extra symmetry to guess the operator  $\hat{h}_E^{(\textmd{opt})}(x)$.

The Eq.(6) represents a purification of probe state at time $t$, where the initial probe state is in the pure state $|\psi\rangle$. Suffering from the Markovian dephasing environment, at time $t$ the non-diagonal term of probe state should be like the Eq.(2): $\rho_{10}(t)=\rho_{10}(0)\exp(-2\gamma t)$ for a single particle. It is easy to verify it by deriving the reduced density matrix in the S space. Obviously, the state in Eq.(6) is the purification of probe state at time $t$ in the Markovian environment. In the similar way, a purification of probe state in the correlated environment can be chosen as shown in Eq.(21) and Eq.(25). It is worth stressing that, the initial state of environment perhaps isn't one of the real environment, due to that it is chosen to purify the state of system for obtaining the QFI.

 \end{document}